\documentclass[12pt]{article}
\usepackage{hyperref}
\usepackage{graphicx}
\usepackage{wrapfig}
\usepackage{lipsum} 

\usepackage{amssymb,amsmath,amsfonts,latexsym,graphicx}
\usepackage[numbers]{natbib}
\usepackage{float} 

\begin{document}

\begin{center}
\Large \bf Multi-Class Abnormality Classification Task in Video Capsule Endoscopy \rm

\vspace{1cm}

\large Dev Rishi Verma, Vibhor Saxena, Dhruv Sharma, Arpan Gupta

\vspace{0.5cm}

\normalsize

Institute of Engineering and Technology, JK Lakshmipat University, Jaipur, 302026 Rajasthan, India

\vspace{5mm}

Email: \texttt{\{devrishiverma, vibhorsaxena, dhruvsharma, arpan.gupta\}@jklu.edu.in}

\vspace{1cm}

\end{center}

\begin{abstract}
In this work for Capsule Vision Challenge 2024, we addressed the challenge of multi-class anomaly classification in video capsule Endoscopy (VCE)\cite{handa2024capsule} with a variety of deep learning models, ranging from custom CNNs to advanced transformer architectures. The purpose is to correctly classify diverse gastrointestinal disorders, which is critical for increasing diagnostic efficiency in clinical settings. We started with a baseline CNN model and improved performance with ResNet\cite{ref:resnet} for better feature extraction, followed by Vision Transformer (ViT)\cite{ref:vit} to capture global dependencies. We further improve the results by using Multiscale Vision Transformer (MViT)\cite{ref:mvitv2} for improved hierarchical feature extraction, while Dual Attention Vision Transformer (DaViT) \cite{ref:davit} delivered best results by combining spatial and channel attention methods. Our best balanced accuracy on validation set \cite{Handa2024training} was \textbf{0.8592} and Mean AUC was \textbf{0.9932}. This methodology enabled us to improve model accuracy across a wide range of criteria, greatly surpassing all other methods.Additionally, our team capsule commandos achieved 7th place ranking with a test set \cite{Handa2024testing}performance of Mean AUC: \textbf{0.7314} and balanced accuracy: \textbf{0.3235}.
\end{abstract}

\section{Introduction}\label{sec1}
The entire gastrointestinal (GI) tract, especially the small intestine, which is difficult to reach with traditional endoscopy, can be examined with Video Capsule Endoscopy (VCE) \cite{Melson2021}, a non-invasive diagnostic technique, whereby a capsule fitted with a camera captures images (or videos) at regular intervals as it passes through the GI tract. The data is captured at a rate ranging from 2 to 35 FPS, having a resolution of around $256 \times 256 \; (H \times W)$. However, manual analysis of this captured data is time-consuming and prone to errors due to the enormous volume of video frames produced by every procedure. Learning-based automated systems have enormous potential to improve the precision and effectiveness of identifying gastrointestinal anomalies such as ulcers, polyps, and bleeding \cite{handa2024capsule}. Recent developments in transformer architectures have demonstrated incredible promise in this area, especially with models such as the Vision Transformer (ViT) \cite{ref:vit}, which use attention mechanisms \cite{Vaswani2017} and hierarchical processing of visual information to dramatically improve performance on image recognition tasks.
 \cite{ref:vit}.

In order to solve this issue, the Capsule Vision 2024 Challenge promotes the creation of sophisticated AI models that can classify abnormalities across multiple classes. The dataset provided consists of images from 3 publicly available datasets, SEE-AI \cite{Yokote2024:SEEAI}, KID \cite{koulaouzidis2017kid}, KVASIR-capsule \cite{Smedsrud2021:KVASIR} and one private AIIMS \cite{Goel2022:AIIMSDataset} dataset. In order to improve the classification performance, we experiment with a variety of deep learning architectures, starting with a custom CNN and working our way up to more complex models like the Vision Transformer (ViT) and its variations. By combining convolutional techniques with attention mechanisms, our method takes advantage of the spatial-temporal complexity of VCE image data. This improves the model's capacity to identify both local and global dependencies within the video frames. Furthermore, prior research has shown that deep learning methods can successfully detect anomalies in VCE, which offers a strong basis for our DaViT \cite{ref:davit} model's performance.

We present a detailed analysis of our experiments and the associated results. In Section~\ref{sec2}, we provide the methods and models used for training, along with the details of data augmentation. Section~\ref{sec3} and \ref{sec4} present our results and related discussions, respectively. Finally, conclusion is given in Section~\ref{sec5}.

\section{Methods}
\label{sec2}

\begin{figure}[ht]
    \centering
    \includegraphics[width=0.5\textwidth]{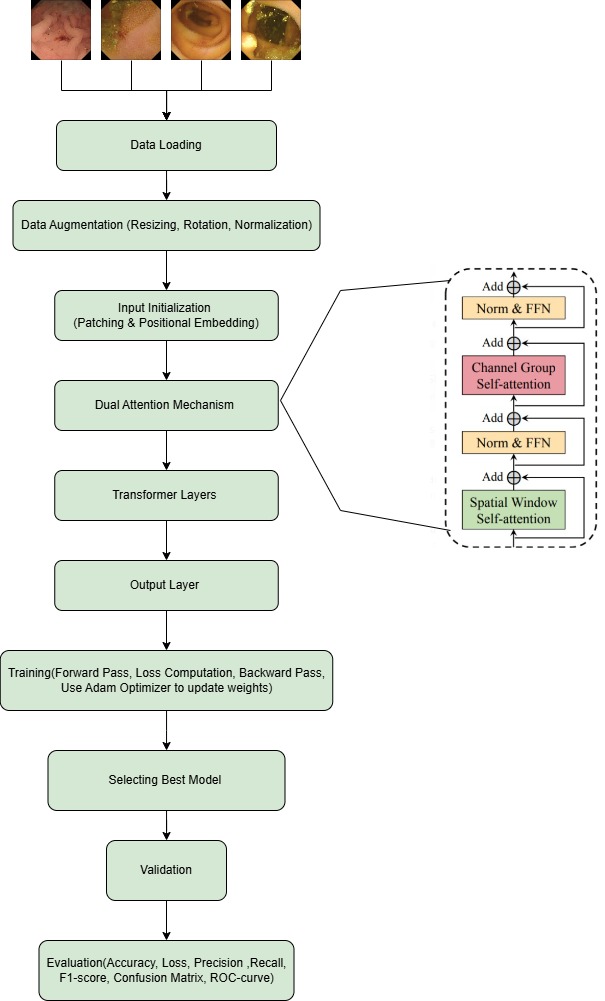}
    \caption{Block diagram of the developed DaViT pipeline. DaViT model, same as in \cite{ref:davit}.}
    \label{fig:davit}
\end{figure}

Figure~\ref{fig:davit} illustrates our experimental setup. The steps and models are described in the subsequent subsections.
\subsection{Preprocessing}
To improve performance in medical imaging, the model's preprocessing involves particular transformations for the training and validation datasets. Images in the training set are resized to 224 × 224 pixels and rotated up to 45 degrees in addition to being randomly flipped horizontally and vertically. These methods aid in enhancing the data and strengthen the model's resistance to scale and orientation changes. In order to stabilize model learning, images are also transformed into tensors and normalized using standard RGB mean and standard deviation values.

\subsection{CNN (Convolutional Neural Network)}

With three convolutional layers that detected spatial information, the first model made use of a Convolutional Neural Network (CNN) designed for structured data, such as photographs. Each layer maintains input dimensions while using a 3x3 kernel, stride of 1, and padding of 1. Complex data associations are captured by the non-linearity introduced by the ReLU activation function.As seen with max pooling layers, the spatial dimensions are cut in half while maintaining the crucial information by employing a 2x2 kernel and stride of 2. This enables the model to concentrate on crucial elements of feature maps for examination.
The number of classes in our dataset corresponds to the projected class scores.

\subsection{ResNet (Residual Network)}

We refined a pretrained ResNet50 model for image classification, known for its deep network architecture and use of residual connections to address the vanishing gradient problem in very deep networks. By allowing some layers to skip steps and maintain a steady flow of gradients, the model can effectively learn even with many layers, improving overall performance. We modified the ResNet code provided by the organizers by changing the input image size to 224x224, enhancing the model's accuracy in capturing fine-grained features and improving classification performance. Our ResNet-50 setup includes an initial layer for basic feature extraction, residual blocks for learning complex patterns with better connections, and a final layer customized for accurate classification based on specific classes in the dataset.

\subsection{ViT (Vision Transformer) }

The Vision Transformer model processes images by breaking them down into patches and reshaping them for processing by the Transformer. A trainable linear projection maps patches to a fixed size, resulting in patch embeddings. Position embeddings are added for positional information. The Transformer encoder consists of layers alternating between self-attention and multi-layer perceptron blocks. ViT has less built-in image-specific knowledge compared to CNNs, learning spatial relationships from scratch. Feature maps from a CNN can also be used as input patches for flexibility in patch size and representation. This approach allows ViT to learn spatial relationships between patches independently from pre-established image features.

\subsection{MViT (Multiscale Vision Transformer)}

The Multiscale Vision Transformer (MViT) architecture utilizes transformer blocks in different stages to process images efficiently. The model, based on the pre-trained mvit v2 large, adjusts to input complexities by operating at various spatial and temporal resolutions. At the core of MViT is the Multi Head Pooling Attention (MHPA) mechanism, which pools input sequences before attention calculations to handle varying resolutions effectively. The pooling operator adjusts sequence length through specified stride and padding, preserving crucial information while reducing computational load. Attention mechanism processes shortened tensors, capturing essential data relationships for effective learning. Parallelization of attention across multiple heads enhances feature learning, enabling MViT to identify diverse characteristics in the data.
\cite{ref:mvitv2}.

\subsection{DaViT (Dual Attention Vision Transformers)}

The Dual Attention Vision Transformers (DaViT) architecture efficiently captures fine details and overall patterns in visual data through four stages. It incorporates dual attention blocks that use spatial window attention to focus on specific regions within small patches of an image for computational efficiency. Additionally, channel group attention enables different channels to interact and gather global information. DaViT offers various configurations, such as DaViT-Tiny, DaViT-Small, and DaViT-Base, to optimize performance according to task complexity and dataset size. This versatility makes DaViT a promising solution for a wide range of vision-related challenges, striking a balance between efficiency and effectiveness in processing visual information.
\cite{ref:davit}.

\section{Results}
\label{sec3}

A number of important metrics, such as the precision-recall curve, ROC curve, and per-class precision, recall, and F1 score, were used to assess the DaViT model's performance on the validation dataset.\cite{Handa2024training}.
These metrics assess the model's overall efficacy across a range of abnormalities and offer insight into how well it can differentiate between classes.Code and implementation for these can be viewed at our github \cite{ref:github}

\begin{figure}[H]
    \centering
    \begin{minipage}{0.45\linewidth}
        \centering
        \includegraphics[width=\linewidth]{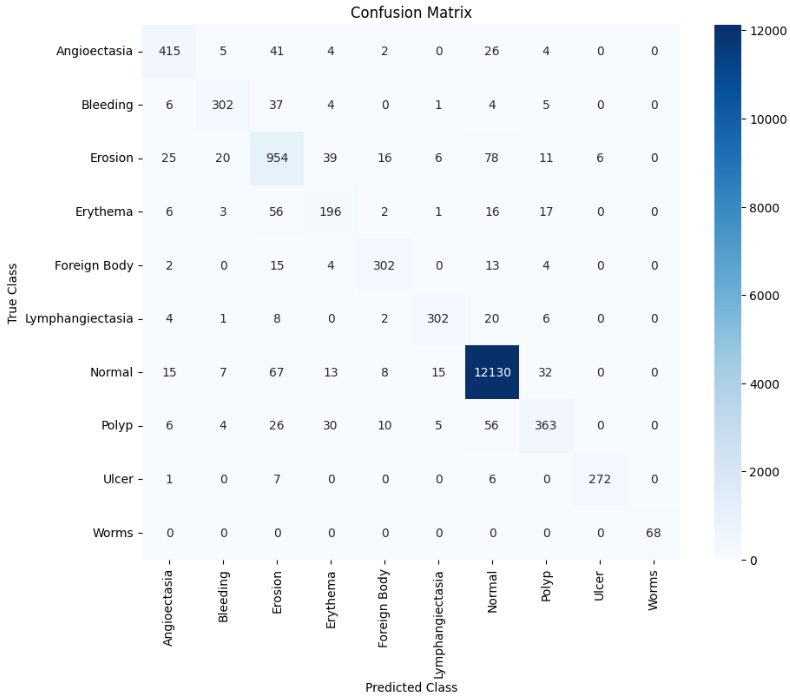}
        \caption{Confusion Matrix for the DaViT Model on the validation set.}
        \label{fig:confusion-matrix}
    \end{minipage}
    \hfill
    \begin{minipage}{0.45\linewidth}
        \centering
        \includegraphics[width=\linewidth]{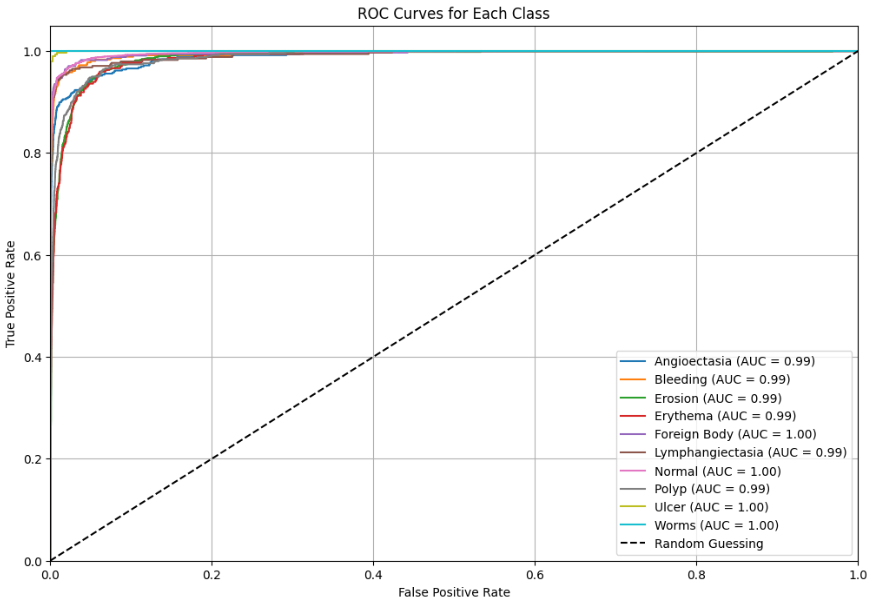}
        \caption{ROC Curve for the DaViT Model on the validation set.}
        \label{fig:roc-curve}
    \end{minipage}
\end{figure}

\label{subsec1}
In our experiments, the DaViT model outperformed traditional methods across key metrics. Table~\ref{table:results} give the classification metrics for the 10 classes when our finetuned DaViT model is evaluated on the validation set (having 16132 samples). The detailed performance results when comparing different models, are shown in Table \ref{table:compare}.

\begin{table}[H]
    \centering
    \begin{tabular}{|c|c|c|c|c|}
        \hline
        \textbf{Classes} & \textbf{Precision} & \textbf{Recall} & \textbf{F1-Score} &  \textbf{Specificity}\\
        \hline
        Angioectasia  & 0.8645 & 0.8350 & 0.8495 & 0.9958\\
        Bleeding & 0.8830 & 0.8412 & 0.8616 & 0.9974\\
        Erosion & 0.7877&0.8259  &0.8064  & 0.9828\\
        Erythema & 0.6758& 0.6599 & 0.6678 &0.9940\\
        Foreign Body & 0.8830& 0.8882 & 0.8856 &0.9974\\
        Lymphangiect & 0.9151 & 0.8804 &0.8974&0.9982\\
        Normal  & 0.9822 &0.9872 & 0.9847&0.9430\\
        Polyp & 0.8212& 0.7260& 0.7707&0.9949\\
        Ulcer & 0.9784& 0.9510& 0.9645&0.9996\\
        Worms & 1& 1& 1&0.990\\
        \hline
    \end{tabular}
    \caption{Classification report for all classes on validation set}
    \label{table:results}
\end{table}

\begin{table}[H]
    \centering
    \begin{tabular}{|p{2cm}|p{2cm}|p{2cm}|p{2cm}|p{2cm}|p{1.5cm}|p{2cm}|}
        \hline
          & \textbf{Mean Specificity} & \textbf{Mean Avg. Precision} & \textbf{Mean Sensitivity} & \textbf{Mean F1-score} & \textbf{Mean AUC} & \textbf{Balanced Acc.} \\
        \hline
        \textbf{Custom CNN} &  0.9642  &  0.6545  &  0.5703  &  0.6265  &  0.9524 & 0.5703 \\
        \hline
        \textbf{ResNet50} &  0.9844  &  0.8685  &  0.7963  &  0.8192  & 0.9900  & 0.7963  \\
        \hline
        \textbf{ViT} &  0.9803  &  0.8206  & 0.7505 & 0.7710 &  0.9847  & 0.7505 \\
        \hline
        \textbf{mViTv2-l} & 0.9890 & 0.9121 & 0.8307 & 0.8505 & 0.9933 & 0.8307 \\
        \hline
        \textbf{DaViT-s} &  \textbf{0.9904}  &  \textbf{0.9147}  &  0.8595  &  \textbf{0.8688}  & \textbf{0.9932} & 0.8595 \\
        \hline
        \textbf{DaViT-s (WRS)} &  0.9900  & 0.8989 & 0.8604 & 0.8600 &  0.9902  & 0.8604 \\
        \hline
    \end{tabular}
    \caption{Performance comparison of models on key metrics.}
    \label{table:compare}
\end{table}
\begin{table}[H]
    \centering
    \begin{tabular}{|l|c|}
        \hline
        \textbf{Metric} & \textbf{Value} \\
        \hline
        Mean AUC & 0.7314 \\
        Balanced Accuracy & 0.3235 \\
        Average Precision & 0.2534 \\
        Average Sensitivity & 0.3235 \\
        Average F1-Score & 0.2274 \\
        Average Specificity & 0.9361 \\
        \hline
    \end{tabular}
    \caption{Test set  results for our Davit model.}
    \label{table:results}
\end{table}

\section{Discussion}
\label{sec4}
This study addressed the multi-class anomaly classification difficulty in Video Capsule Endoscopy (VCE), with a focus on transformer model performance. While our early models, the custom CNN and ResNet50, offered a solid platform for analysis (accuracies of \textbf{0.8502} and \textbf{0.9308}, respectively), ResNet50 outperformed the ViT model, which achieved an accuracy of \textbf{0.9138}. MViT increased our results to \textbf{0.9440}, demonstrating the transformer designs' ability to handle complicated data distributions and extract the nuanced patterns required for accurate anomaly identification.
DaViT small outperformed all other models, with an outstanding accuracy of \textbf{0.9487}. We applied  weighted random sampling (WRS) to our DaVit model which showed an accuracy of \textbf{0.9433}, showing persistently high performance.
The evaluation metrics for transformer models provide more information about their capabilities. MViT had a mean specificity of 0.9890, a mean F1-score of 0.9933, and a balanced accuracy of 0.8307, indicating that it can effectively detect both positive and negative cases. Similarly, DaViT Small achieved a mean AUC of 0.9932 and a balanced accuracy of 0.8595, demonstrating its skill in discriminating between distinct classes within the dataset.
On the test set, DaViT Small achieved a mean AUC of \textbf{0.7314}, a balanced accuracy of \textbf{0.3235}, and an average precision of \textbf{0.2534}. These results highlight the model's ability to generalize its performance, though the gap between validation and test metrics suggests further optimization and dataset diversity may be needed to improve test-time accuracy. The model also demonstrated a mean sensitivity of \textbf{0.3235}, mean F1-score of \textbf{0.2274}, and mean specificity of \textbf{0.9361} on the test set, confirming its capability in identifying specific anomalies but also pointing out areas for improvement in handling class imbalances and challenging cases.
Interestingly, the addition of WRS did not result in significant performance improvements for DaViT small, implying that the model's intrinsic strengths and architectural design are critical for classification success.
Finally, the findings of this study show the transformative potential of complex transformer models in medical diagnostics. DaViT Small's high performance verifies its suitability for the Capsule Vision 2024 challenge and demonstrates its ability to improve accuracy in automated diagnoses, resulting in better patient outcomes and operational efficiency in healthcare. Future research should concentrate on these designs and their use in a range of diagnostic contexts in order to achieve even higher accuracy and dependability.

\section{Conclusion}
\label{sec5}
In conclusion, our study on multi-class anomaly classification in video capsule endoscopy (VCE) demonstrates the potential of advanced transformer-based architectures, especially the DaViT model, to achieve excellent accuracy and robustness. By methodically examining a number of models, such as a custom CNN, ResNet50, Vision Transformer (ViT), MViT, and DaViT, we found that while simpler architectures like CNNs and ResNet provided foundational insights, they fell short in handling intricate visual patterns in medical data.

DaViT emerged as the best-performing model with an accuracy of \textbf{0.9466} on validation data, alongside notable gains in recall, precision, F1-score, and AUC metrics. On the test set for the Capsule Vision 2024 challenge, DaViT achieved a mean AUC of \textbf{0.7314}, a balanced accuracy of \textbf{0.3235}, and an average precision of \textbf{0.2534}. These findings underscore the model's ability to generalize but also highlight the challenges posed by real-world datasets, such as class imbalances and diverse visual patterns.

Our team's efforts culminated in securing the \textbf{7th rank} in the Capsule Vision 2024 challenge, showcasing the effectiveness of DaViT in addressing multi-class anomaly classification in Video Capsule Endoscopy (VCE). DaViT's ability to efficiently capture both local and global information makes it a compelling choice for medical image classification tasks.

These findings underline the transformational potential of advanced transformer models in medical diagnostics. Future research should focus on refining these architectures, incorporating more diverse datasets, and addressing the challenges identified in test performance to further enhance diagnostic accuracy and robustness in real-world applications.

\section{Acknowledgments}
\label{sec6}
As participants in the Capsule Vision 2024 Challenge, we fully comply with the competition's rules as outlined in \cite{handa2024capsule}. Our AI model development is based exclusively on the datasets provided in the official release in \cite{handa2024capsule}.

\bibliographystyle{unsrtnat}
\bibliography{sample}

\end{document}